# Thermal Control of Transmission Spectral Characteristics in a Ring Resonator

Nikolai N. Klimov[1,2], Mario C. M. M. Souza[3], Zeeshan Ahmed[1*]

[1]Thermodynamic Metrology Group, Sensor Science Division, Physical Measurement Laboratory, National Institute of Standards and Technology, Gaithersburg, MD, USA

[2] Joint Quantum Institute, University of Maryland, College Park, MD, USA

[3] Gleb Wataghin Physics Institute, University of Campinas, Campinas, SP, Brazil

*Corresponding author: zeeshan.ahmed@nist.gov

We report on our observation of thermal control of transmission spectra of a micro-ring resonator coupled to a low finesse etalon. Thermal tuning of the device results in relative changes in phase between ring and etalon modes along with changes in coupling strength and losses between the etalon and ring resonator resulting in a wide variety of line shapes ranging from a symmetric to asymmetric Fano-like to electromagnetically induced transparency (EIT)-like behavior. The capability to modify spectral line shapes by thermally tuning the coupling strength and losses between two resonators has obvious applications in optical communication networks and relevance to sensor applications where it may be exploited for developing novel cutoff switches and has significant, deleterious effect on measurement uncertainties of large dynamic range sensors.

The past two decades have witnessed tremendous advances in photonics leading to the development of novel photonic devices such as Bragg mirrors, ring resonators and photonic crystal cavities.[1-3] At its simplest a ring resonator is a channel-dropping filter where only the wavelength of light that satisfies the cavity's resonant condition is dropped. The resonant condition is directly proportional to the refractive index of the waveguide and to a lesser extent the surrounding environment.[2, 4-7]

In recent years, this dependence of the resonance wavelength on modal index has been exploited in a wide range of sensing applications ranging from temperature[5, 8, 9], pressure[10] to chemical sensing in both the gas and liquid phase.[11] Devices from the broader whispering gallery mode resonator family have shown wide variety of behaviors such as polarization mode coupling,[12, 13] avoided mode crossings[13], "synthesis" of photonic molecules[14] and EIT-like transparency window[15-21]. There is increasing interest in exploiting such phenomenon to either achieve higher sensitivities in threshold sensor configurations or to develop novel re-configurable communication switches.[18, 21] Less well understood is the impact of such transition on measurement uncertainties of photonic sensors.

In this study, we present our observation of thermally driven continuous conversion of spectral line shapes from symmetric to asymmetric Fano-like to EIT-like features and discuss the impact of such lineshape changes in sensor performance . The device consists of a Si straight-probe waveguide with cross-section of 610 nm × 220 nm, evanescently coupled to a 11 µm diameter ring over an air gap of 130 nm and covered with a 1.5 µm thick Plasma-enhanced chemical vapor deposition (PECVD) $SiO_2$. The silicon photonic sensors were made at CEA-LETI (Laboratoire d'Electronique et de Technologie de l'Information, France)[1] fabrication facility using standard CMOS compatible manufacturing technology.

The device is probed using a custom built interrogation system that has been described in detail elsewhere[5]. In this setup a C-band laser (New Focus TLB-6700 series) is swept over the sensor resonance. Ten percent of laser power was immediately picked up from the laser output for wavelength monitoring (HighFinesse WS/7) while the rest, after passing through polarization paddles is injected into the photonic device and detected by a large sensing-area power meter (Newport, model 1936-R). Light was coupled into and out of the waveguide using grating couplers to which single mode fibers were bonded using UV-curable epoxy. Unless noted otherwise, all experiments were performed with ≈120 nW of incident laser power.

At 25 °C the ring resonator shows modes located at ≈1524.06 nm, 1532.30 nm, 1540.65, 1549.08 nm and 1557.59 with free spectral range (FSR) of ≈ 8 nm indicating $n_{eff}$ of 4.297. We observe a wide variety of lineshapes including symmetric (mode 3 and 5), asymmetric Fano like (mode 1 and 4) and an EIT-like lineshape (mode 2). The symmetric lineshape modes show a clear splitting of the resonance. It is well known that such mode splitting can arise from loss off symmetry due to surface roughness induced coupling to backward propagating mode.[22, 23] For the 1540 nm mode (Fig 1), the splitting is 6.25 GHz (≈50 pm). Consistent with previous steps the ring resonance shows ≈75 pm/K shift in peak position as the temperature is varied.

---

[1] Disclaimer: Certain commercial fabrication facility, equipment, materials or computational software are identified in this paper in order to specify device fabrication, the experimental procedure and data analysis adequately. Such identification is not intended to imply endorsement by the National Institute of Standards and Technology, nor is it intended to imply that the facility, equipment, material or software identified are necessarily the best available.

Thermal cycling 0.25 °C steps, over the range of -5 °C to 45 °C, we observe a series of periodic changes in line shape of each of the modes every 10 °C (at 14 °C, 24 °C and 34 °C for 1540 nm mode) that reproduce the line shape changes observed in Fig 1a (Fig 2). The transition region is only 3 °C wide (225 pm) and coincides with ring resonance crossing over the trough in the background oscillation of the etalon. The half-period of the background oscillation is ≈750 pm, thus every 10 °C the ring mode is tuned from trough to trough. As shown in Fig 1B, the background oscillation has a FSR of 9.8 nm. We note that neither the background nor the ring resonance show any anomalous dependence on incident laser power which allows us to rule out any higher order effects such as Autler-Townes splitting where the splitting changes with pump power[18]. Furthermore, we do not observe any spectral changes when the input or output polarization of light is flipped, suggesting polarization mode crossing can be ruled out (data not shown).

Our results are consistent with transition caused by interference between a high-finesse ring resonance and a low finesse background resonance that results in Fano and EIT like transmission characteristics. Such transitions have previously been observed for a Fabry-Perot (FP) resonator coupled to micro-ring[24] or disk[25] resonator device where the transition was driven by slight changes in resonator losses and coupling strength between the resonator. The coupling strength was tuned by changing the air-gap between the waveguide and ring resonator[24, 25]. Theoretical modeling of the devices indicated that in the over- and under-coupled regime, changes in coupling strength result in gradual change in the phase of light stored in the ring resulting in a transition from constructive to destructive transition between the etalon and ring resonator [24, 25].

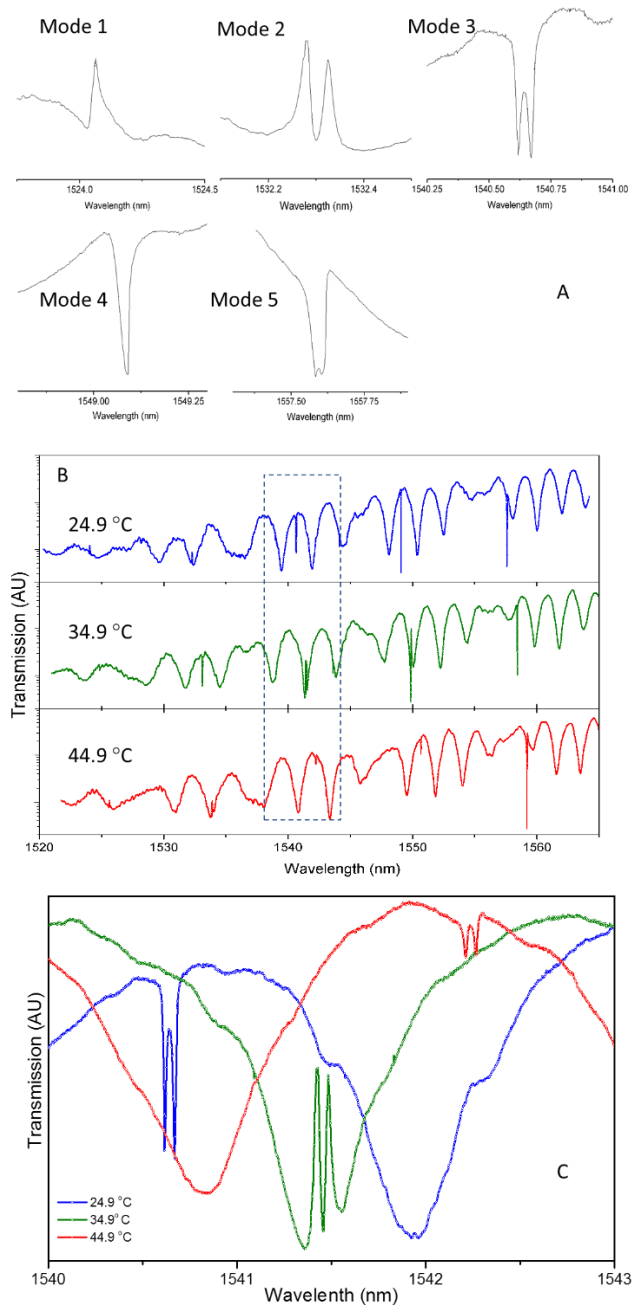

Figure 1: a) Ring modes in the 1520 nm to 1570 nm region show a diverse array of lineshapes. b) Temperature depedent spectra highlight changes in line extinction and lineshape (c) as the ring resonance is thermally detuned from the FP resonance

In our device, spectral line shape changes are achieved by means of thermal tuning rather than an adjustment of air-gap. It is well known that as resonance redshifts (longer wavelength) coupling strength increases and finesse decreases [26]. The extinction ratio will decrease with increasing wavelength when resonator is over-coupled; in the case of under-coupled resonator the extinction ratio will increase until critical coupling is achieved and then decrease [26]. This effect is readily apparent in

Fig 1 where we observe the extinction ratio (depth of resonance line) grow and line shape change significantly with increasing wavelength. The coupled mode model (see ref 24) can therefore be generally applied to explain the observed periodic changes in line shape as arising due to thermally induced detuning of the ring mode from the FP mode. This simplified picture however, falls short on two accounts: a) coupling strength does not appear to change monotonically with temperature as evidenced by variation in extinction observed with temperature (see Fig 1b) b) the off-resonance line shapes differs significantly from the predicted EIT like line shape.

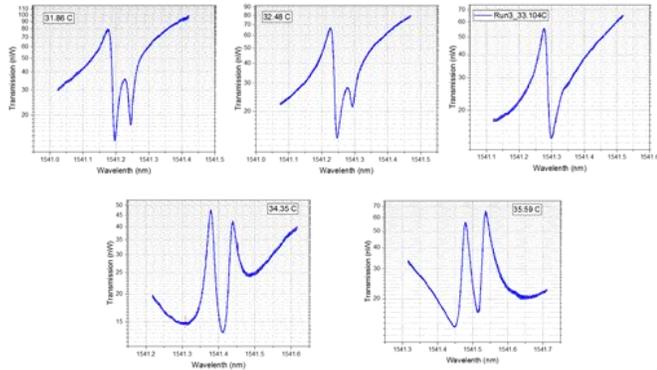

**Figure 2: Temperature tuning reproduces line shapes observed in Fig 1**

As shown in Fig 2 on-resonance the 1541 nm mode shows a symmetric profile with strong extinction that changes to asymmetric Fano profile when detuned. Off-resonance the line shape resembles an emission peak. As shown in Fig 3, modeling of the ring resonator-FP system suggests off-resonance spectra exhibit a symmetric absorption like profile in the under and critical coupling regime; in the over coupled regime the line shape change to a more complex EIT like profile. The observed changes in experimental spectra suggest the coupling strength varies periodically, increasing when off-resonance and decreasing when detuned or on-resonance. The periodicity in coupling strength cannot be explained solely by a monotonic thermally induced change in coupling strength.

    We hypothesize that the shortfalls in the model arise due to non-zero coupling between the counter-clockwise (CCW) and clockwise (CW) mode in the ring resonator. Due to this coupling condition, the relative phase of the CCW and CW modes relative to the FP modes is critical as small changes in this phase can drive large changes in line shape. We model these change in phase by computing the spectra of coupled device that was systematically displaced laterally with respect to the FP resonator (air-gap is invariant). As shown in Fig 4 a small relative mismatch in phase between FP and ring modes is sufficient to generate

the wide changes in line shape and extinction similar to those observed experimentally.

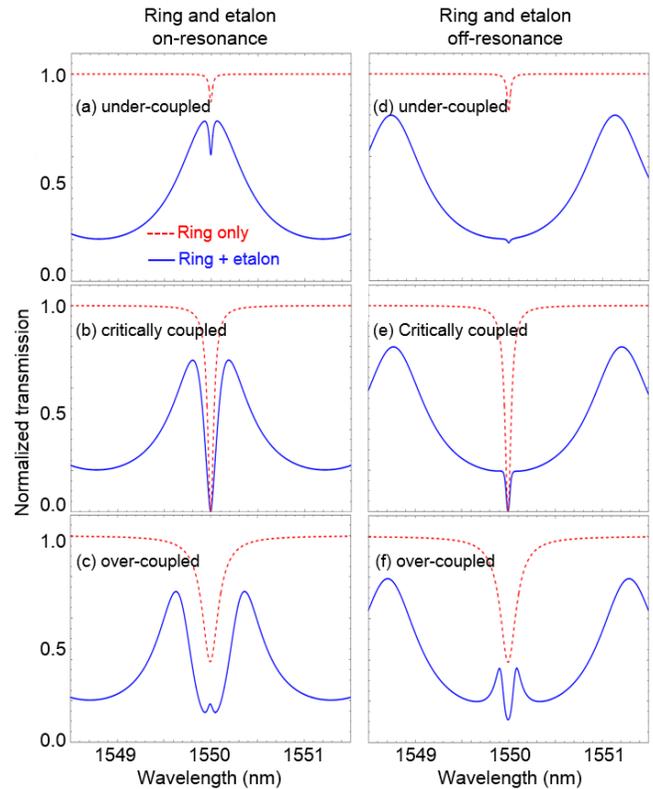

Figure 3: Calculated transmission spectra of a microring resonator embedded in a Fabry-Perot etalon (blue traces) at different coupling conditions. The transmission spectra of the microring alone, i.e. if the etalon effect was not present, are presented as red-dashed traces.

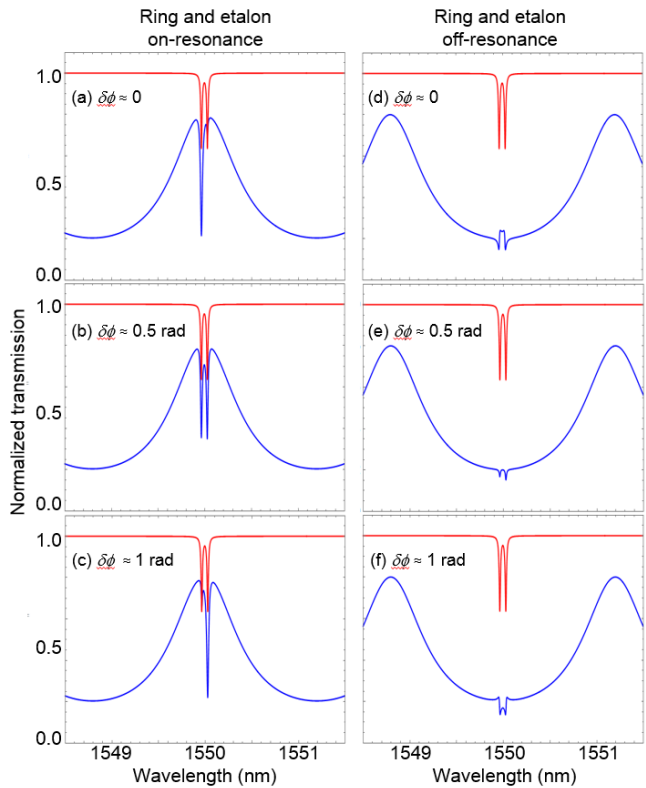

Figure 4: Transmission spectra of a microring resonator with coupling between propagative and counter-propagative modes embedded in a Fabry-Perot etalon (blue traces) for different phases between ring and etalon. The transmission spectra of the microring alone are presented as red-dashed traces.

As noted above, the ability to manipulate line shape is of great interest to those developing optical devices for applications in neuromorphic computing and novel threshold sensors. In the realm of photonic sensor metrology e.g., thermometry, such phenomenon can prove to be detrimental. In the present case the device under study shows a linear temperature dependence of 75 pm/K with quality factors of 90000. Such a device can resolve µK deviations in temperature[5] with accuracy on the order of tens of mK. However, when using the wavelength-swept method, the line shape changes complicate any attempt at accurately locating the peak center resulting in uncertainties of ≈575 mK ($k = 1$). If laser locking techniques are employed to track temperature induced changes in wavelength, the lineshape transition can complicate servo's effort to maintain lock on chosen mode (CCW or CW). Mode hoping between two modes can result in off-set errors of ≈649 mK ($k = 1$). Effectively this limits the use of such a device over narrow temperature window(s) where the superior temperature resolution of these devices could put to effective use. The uncertainties over broader operating ranges can be achieved (or surpassed) by more rudimentary devices such as a Bragg grating[7, 9] nullifying the advantages of high Q-factors devices.

Our results highlight the need for development of standardized criteria used to characterize and select device suitable for different sensing applications. Devices such as the once described here maybe sufficient for application that require limited dynamic range e.g. threshold detection. Their use in application such as thermometry or manometry which often requires large operating ranges is likely to lead to inconsistent results. For such applications, mode conversion or interactions need to be minimized if are to achieve measurement accuracies on par with legacy technology.

### References:

1. P. Yeh and A. Yariv, "Bragg reflection waveguides," Optics Communications **19**, 427-430 1976.
2. Y. Xu, Y. Li, R. K. Lee, and A. Yariv, "Scattering-theory analysis of waveguide-resonator coupling," Physical Review E **62**, 7389-7404 2000.
3. Q. Quan and M. Loncar, "Deterministic design of wavelength scale, ultra-high Q photonic crystal nanobeam cavities," Optics Express **19**, 18529-18542 2011.
4. W. W. Rigrod, "The Optical Ring Resonator," The Bell System Technical Journal, 907-916 1965.
5. H. Xu, M. Hafezi, J. Fan, J. Taylor, G. F. Strouse, and Z. Ahmed, "Ultra-Sensitive Chip-Based Photonic Temperature Sensor Using Ring Resonator Structures," Optics Express **22**, 3098-3104 2014.
6. M. Á. Guillén-Torres, K. Murray, H. Yun, M. Caverley, E. Cretu, L. Chrostowski, and N. A. F. Jaeger, "Effects of backscattering in high-Q, large-area silicon-on-insulator ring resonators," Optics Letters **41**, 1538-1541 2016.
7. Z. Ahmed, J. Filla, W. Guthrie, and J. Quintavall, "Fiber Bragg Gratings Based Thermometry," NCSLI Measure **10**, 24-27 2015.
8. N. N. Klimov, M. Berger, and Z. Ahmed, "Characterization of Ring Resonator Structures for Applications in Photonic

9. N. N. Klimov, T. Purdy, and Z. Ahmed, "On-Chip silicon photonic thermometers: from waveguide Bragg grating to ring resonators sensors," in *SPIE Sensing Technology+Applications*, 2015), 948609.
10. X. Zhao, J. M. Tsai, H. Cai, X. M. Ji, J. Zhou, M. H. Bao, Y. P. Huang, D. L. Kwong, and A. Q. Liu, "A nano-opto-mechanical pressure sensor via ring resonator," Opt. Express **20**, 8535-8542 2012.
11. M.-S. Kwon and W. H. Steier, "Microring-resonator-based sensor measuring both the concentration and temperature of a solution," Opt. Express **16**, 9372-9377 2008.
12. M. K. Chin, "Polarization dependence in waveguide-coupled micro-resonators," Optics Express **11**, 1724-1730 2003.
13. S. Ramelow, A. Farsi, S. Clemmen, J. S. Levy, A. R. Johnson, Y. Okawachi, M. R. E. Lamont, M. Lipson, and A. L. Gaeta, "Strong polarization mode coupling in microresonators," Optics Letters **39**, 5134-5137 2014.
14. B. Peng, Ş. K. Özdemir, J. Zhu, and L. Yang, "Photonic molecules formed by coupled hybrid resonators," Optics Letters **37**, 3435-3437 2012.
15. Y. Yang, S. Saurabh, J. Ward, and S. N. Chormaic, "Coupled-mode-induced transparency in aerostatically tuned microbubble whispering-gallery resonators," Optics Letters **40**, 1834-1837 2015.
16. R. Henze, T. Seifert, J. Ward, and O. Benson, "Tuning whispering gallery modes using internal aerostatic pressure," Optics Letters **36**, 4536-4538 2011.
17. X. Jin, Y. Dong, and K. Wang, "Stable controlling of electromagnetically induced transparency-like in a single quasi-cylindrical microresonator," Optics Express **24**, 29773-29780 2016.
18. B. Peng, Ş. K. Özdemir, W. Chen, F. Nori, and L. Yang, "What is and what is not electromagnetically induced transparency in whispering-gallery microcavities," **5**, 5082 2014.
19. Y. Zheng, J. Yang, Z. Shen, J. Cao, X. Chen, X. Liang, and W. Wan, "Optically induced transparency in a micro-cavity," Light Sci Appl. **5**, e16072 2016.
20. A. Naweed, G. Farca, S. I. Shopova, and A. T. Rosenberger, "Induced transparency and absorption in coupled whispering-gallery microresonators," Physical Review A **71**, 043804 2005.
21. Y. Zhang, T. Mei, and D. H. Zhang, "Temporal coupled-mode theory of ring-bus-ring Mach-Zehnder interferometer," Applied Optics **51**, 504-508 2012.
22. B. E. Little, J.-P. Laine, and S. T. Chu, "Surface-roughness-induced contradirectional coupling in ring and disk resonators," Optics Letters **22**, 4-6 1997.
23. M. Borselli, K. Srinivasan, P. E. Barclay, O. Painter, V. D. W., I. V. S., M. H., S. E. W., and K. H. J., "Rayleigh scattering, mode coupling, and optical loss in silicon microdisks," Applied Physics Letters **85**, 3693-3695 2004.
24. Z. Zhang, G. I. Ng, T. Hu, H. Qiu, X. Guo, W. Wang, M. S. Rouifed, C. Liu, and H. Wang, "Conversion between EIT and Fano spectra in a microring-Bragg grating coupled-resonator system," Applied Physics Letters **111**, 081105 2017.
25. W. Liang, L. Yang, J. K. Poon, Y. Huang, K. J. Vahala, and A. Yariv, "Transmission characteristics of a Fabry-Perot etalon-microtoroid resonator coupled system," Optics Letters **31**, 510-512 2006.
26. W. R. McKinnon, D. X. Xu, C. Storey, E. Post, A. Densmore, A. Delâge, P. Waldron, J. H. Schmid, and S. Janz, "Extracting coupling and loss coefficients from a ring resonator," Optics Express **17**, 18971-18982 2009.